\documentclass[%
superscriptaddress,
nofootinbib,
nobibnotes,
amsmath,amssymb,
aps,
floatfix,
]{revtex4-2}

\usepackage{placeins}
\usepackage{booktabs}
\usepackage{dcolumn}
\usepackage{array}
\usepackage{longtable}
\usepackage{float}
\usepackage{hyphenat}
\usepackage{natbib}
\usepackage{hyperref}

\usepackage{amsmath,amsfonts,amsthm,amssymb}
\usepackage{bm,graphicx,graphics,color}
\usepackage{epsf,epsfig}

\def\bx{\mathbf{x}}

\def\bv{\mathbf{v}}

\def\bk{\mathbf{k}}

\def\no{\nonumber}
\def\lb{\label}
\def\be{\begin{equation}}
\def\ee#1{\label{#1}\end{equation}}
\newcommand{\ben}{\begin{eqnarray}}
\newcommand{\een}{\end{eqnarray}}

\begin{document}

\title{Jeans Instability from post-Newtonian Boltzmann equation}

\author{Gilberto M. Kremer}
\email{kremer@fisica.ufpr.br}
\affiliation{Departamento de F\'{i}sica, Universidade Federal do Paran\'{a}, Curitiba 81531-980, Brazil}

\begin{abstract}
Jeans instability within the framework of post-Newtonian Boltzmann and Poisson equations are analyzed. The components of the energy-momentum tensor are calculated from a post-Newtonian Maxwell-J\"uttner distribution function.  The perturbations of the distribution function and gravitational potentials  from their background states with the representation of the perturbations as plane waves  lead to a dispersion relation with post-Newtonian corrections. The influence of the post-Newtonian approximation on the Jeans mass is determined and it was shown that the mass necessary for an overdensity to begin the gravitational collapse in the post-Newtonian theory is smaller than the one in the Newtonian theory.
\end{abstract}

\maketitle

\section{Introduction}

The first attempt to describe instabilities of self-gravitating fluids from the hydrodynamic equations coupled with the Newtonian Poisson equation   was due to Jeans \cite{Jeans}. He determined from a dispersion relation a wavelength cutoff, nowadays known as Jeans wavelength, such that for small wavelengths than the Jeans wavelength the perturbations propagate as harmonic waves in time but for large  wavelengths the perturbations will grow or decay with time. The  gravitational collapse of  self-gravitating interstellar gas clouds associated with the mass density perturbations which grow exponentially with time is known as Jeans instability \cite{Wein,Coles,BT1}. Physically the collapse of a mass density inhomogeneity occurs whenever the inwards gravitational force is bigger than the outwards pressure force.  

Another method to analyse the Jeans instability is to consider  the collisionless Boltzmann equation coupled with the  Newtonian Poisson equation (see e.g \cite{b1,b2,b3,b5,b6,b7,b8,b9}).

Recently  the Jeans instability  was examined within the framework of the first post-Newtonian theory by considering the  hydrodynamic equations and the Poisson equations  which follow from  this theory \cite{NKR}.

The aim of the present work is to analyse the Jeans instability from the post-Newtonian collisionless Boltzmann equation coupled with the  post-Newtonian Poisson equations.  Apart from the Newtonian gravitational potential in the first post-Newtonian theory appear two more gravitational potentials which are associated with two new  Poisson equations. One of the gravitational potentials is a scalar while the other is a vector  \cite{Wein,Ch1}.

Here the components of the energy-momentum tensor, which appear in the post-Newtonian Poisson equations, are functions of the one-particle distribution function, so that the Poisson equations together with the post-Newtonian Boltzmann equation become a closed system of algebraic equations for the perturbed gravitational potentials. From the solution of the  system of algebraic equations a dispersion relation emerges that is used to determine the influence of the post-Newtonian approximation in the Jeans mass, which is related with the minimum mass necessary for an overdensity to initiate the gravitational collapse.

The paper is structured as follows. In Section \ref{sec2} the first post-Newtonian expressions for the Boltzmann and  Poisson equations  and the equilibrium Maxwell-J\"uttner distribution function are introduced.  The Jeans instability is analyzed in Section \ref{sec3} where perturbations in the background states of the  distribution function and gravitational potentials are considered. The  representation of the perturbations as plane waves results into a dispersion relation where the post-Newtonian influence in the Jeans mass  is analyzed. In Section \ref{sec4} a summary of the results is discussed.

\section{Boltzmann equation}\lb{sec2}

In the first post-Newtonian approximation the components of the metric tensor $g_{\mu\nu}$ reads \cite{Ch1}
\ben
g_{00}=1-\frac{2U}{c^2}+\frac2{c^4}\left(U^2-2\Phi\right),\qquad g_{0i}=\frac{\Pi_i}{c^3},\qquad g_{ij}=-\left(1+\frac{2U}{c^2}\right)\delta_{ij},
\een
where the gravitational potentials  $U$, $\Phi$ and  $\Pi_i$  are given by the Poisson equations  
\ben\lb{jpn2a}
&&\nabla^2U=-\frac{4\pi G}{c^2} \,{\buildrel\!\!\!\! _{0} \over{T^{00}}},
\qquad
 \nabla^2\Phi=-2\pi G \left({\buildrel\!\!\!\! _{2} \over{T^{00}}}+{\buildrel\!\!\!\! _{2} \over{T^{ii}}}\right),
\\\lb{jpn2b}
&&\nabla^2\Pi^i=-\frac{16\pi G}{c}{\buildrel\!\!\!\! _{1} \over{T^{0i}}}+\frac{\partial^2U}{\partial t\partial x^i}.
\een
In the above equations  the energy-momentum tensor is split in orders of $1/c^n$ denoted by ${\buildrel\!\!\!\! _{n} \over{T^{\mu\nu}}}$.

The first post-Newtonian approximation of the Boltzmann equation  written in terms of the Chandrasekhar gravitational potentials  $U$, $\Phi$ and  $\Pi_i$ reads 
\ben\no
\bigg[\frac{\partial f}{\partial t}+v_i\frac{\partial f}{\partial x^i}+\frac{\partial f}{\partial v_i}\frac{\partial U}{\partial x^i}\bigg]\bigg[1+\frac1{c^2}\left(\frac{v^2}2+U\right)\bigg]
+\frac1{c^2}\frac{\partial f}{\partial v_i}\bigg\{v_j\bigg(\frac{\partial\Pi_i}{\partial x^j}
-\frac{\partial\Pi_j}{\partial x^i}\bigg)-2v_i\frac{\partial U}{\partial t}-2\frac{\partial\left(U^2-\Phi\right)}{\partial x^i}
\\\lb{jpn1}
+\frac{\partial\Pi_i}{\partial t}-2v_iv_j\frac{\partial U}{\partial x^j}+v^2\frac{\partial U}{\partial x^i}-v_i
\bigg[\frac{\partial U}{\partial t}+2v_j\frac{\partial U}{\partial x^j}\bigg]
\bigg\}=0.
\een

In kinetic theory of gases the energy-momentum tensor is defined in terms of the one-particle distribution function by \cite{CK}
\ben\lb{jpn3a}
T^{\mu\nu}=m^4c\int u^\mu u^\nu f\frac{\sqrt{-g}\,d^3 u}{u_0}.
\een
Here $u^\mu$ is the particle four-velocity whose components in the first post-Newtonian approximation are
\ben\lb{jpn3b}
u^0=c\left[1+\frac1{c^2}\left(\frac{v^2}2+U\right)\right],\qquad u^i=\frac{u^0v^i}c.
\een

The one-particle distribution function at equilibrium for a relativistic gas is given by the Maxwell-J\"uttner distribution function. Its expression in the first post-Newtonian approximation  in a stationary equilibrium  background where the hydrodynamic velocity vanishes is \cite{KRW}
\ben\lb{jpn4a}
f_{MJ}=f_0
\left\{1-\frac{\sigma^2}{c^2}\left[\frac{15}8+\frac{3v^4}{8\sigma^4}
+\frac{2U v^2}{\sigma^4}\right]\right\},\\ f_0=\frac{\rho_0}{m^4(2\pi \sigma^2)^\frac32}e^{-\frac{v^2}{2\sigma^2}}.
\een
Here $f_0$ denotes the Maxwellian distribution function which is given in terms of the gas particle velocity $\bf v$, the mass density $\rho_0$ and the dispersion velocity  $\sigma=\sqrt{kT_0/m}$. Furthermore, $k$  denotes the Boltzmann constant, $T_0$ the temperature  and $m$ the rest mass of a gas particle. The mass density $\rho_0$ and the temperature $T_0$ are considered to be constants, since they refer to a stationary equilibrium background. The factor $1/m^4$ in the Maxwell-J\"uttner distribution function is due to the fact that it is given in terms of the momentum  four-vector $p^\mu$.

The  first post-Newtonian approximation of the invariant integration element of the energy-momentum tensor (\ref{jpn3a}) is given by \cite{KRW}
\ben\lb{jpn4b}
\frac{\sqrt{-g}\, d^3 u}{u_0}=
\left\{1+\frac1{c^2}\left[2v^2+6U\right]\right\}\frac{d^3v}c.
\een

An equivalent expression for the first post-Newtonian Boltzmann equation (\ref{jpn1}) is obtained from its multiplication by  $[1-({v^2}/2+U)/c^2]$
and by considering terms up to the $1/c^2$ order, yielding
\ben\lb{jpn5}
\frac{\partial f}{\partial t}+v_i\frac{\partial f}{\partial x^i}+\frac{\partial U}{\partial x^i}\frac{\partial f}{\partial v^i}
+\frac1{c^2}\bigg[\left(v^2-4U\right)\frac{\partial U}{\partial x^i}-4v_iv_j\frac{\partial U}{\partial x^j}
-3v_i\frac{\partial U}{\partial t}
+2\frac{\partial \Phi}{\partial x^i}+\frac{\partial \Pi_i}{\partial t}+v_j\left(\frac{\partial \Pi_i}{\partial x^j}-\frac{\partial \Pi_j}{\partial x^i}\right)\bigg]\frac{\partial f}{\partial v^i}=0.
\een

\section{Jeans instability}\label{sec3}

For the analysis of Jeans instability we shall rely on  the Poisson equations (\ref{jpn2a}) and (\ref{jpn2b}) coupled with the Boltzmann equation (\ref{jpn5}).

We begin by writing the gravitational potentials and the one-particle distribution function  as a sum of background and  perturbed terms. The background terms refer to an equilibrium state and are denoted by the subscript zero, while the perturbed terms by the subscript 1. Hence, we write
\ben\lb{jpn6a}
&&f(\bx,\bv,t)=f_{MJ}(\bx,\bv,t)+\epsilon f_1(\bx,\bv,t),
\\ \lb{jpn6b}
&&U(\bx,\bv,t)=U_0(\bx)+\epsilon U_1(\bx,\bv,t),
\\\lb{jpn6c}
&&\Phi(\bx,\bv,t)=\Phi_0(\bx)+\epsilon \Phi_1(\bx,\bv,t),
\\\lb{jpn6d}
&&\Pi_i(\bx,\bv,t)=\Pi_i^0(\bx)+\epsilon\Pi_i^1(\bx,\bv,t).
\een
Above we introduced a small parameter $\epsilon$ which controls that only linear terms in this parameter should be considered. Later on this parameter will be set equal to one.

If we introduce  the representations  (\ref{jpn6a}) --  (\ref{jpn6d}) into the Boltzmann equation  (\ref{jpn5}) and  equate the terms of the same $\epsilon$-order we obtain the following hierarchy of equations
\ben\lb{jpn7a}
\frac{\partial U_0}{\partial x^i}\frac{\partial f_{MJ}^0}{\partial v^i}-\frac{2v^2f_0}{\sigma^2c^2}v_i\frac{\partial U_0}{\partial x^i}
+\frac1{c^2}\Bigg[\left(v^2-4U_0\right)\frac{\partial U_0}{\partial x^i}
-4v_iv_j\frac{\partial U_0}{\partial x^j}+2\frac{\partial \Phi_0}{\partial x^i}+v_j\left(\frac{\partial \Pi^0_i}{\partial x^j}-\frac{\partial \Pi^0_j}{\partial x^i}\right)\Bigg]\frac{\partial f_0}{\partial v^i}=0,
\een
\ben\no
\frac{\partial f_1}{\partial t}+v_i\frac{\partial f_1}{\partial x^i}+\frac{\partial U_1}{\partial x^i}\frac{\partial f_{MJ}^0}{\partial v^i}-\frac{2v^2f_0}{\sigma^2c^2}\left(\frac{\partial U_1}{\partial t}+v_i\frac{\partial U_1}{\partial x^i}\right)
+\frac{\partial U_0}{\partial x^i}\frac{\partial f_1}{\partial v_i}-\frac{4v_iU_1}{c^2\sigma^2}\frac{\partial U_0}{\partial x^i}+\frac1{c^2}\Bigg[\left(v^2-4U_0\right)\frac{\partial U_0}{\partial x^i}+2\frac{\partial \Phi_0}{\partial x^i}
\\\no
-4v_iv_j\frac{\partial U_0}{\partial x^j}+v_j\left(\frac{\partial \Pi^0_i}{\partial x^j}-\frac{\partial \Pi^0_j}{\partial x^i}\right)\Bigg]\frac{\partial f_1}{\partial v^i}+\frac1{c^2}\Bigg[2\frac{\partial \Phi_1}{\partial x^i}
+\frac{\partial \Pi^1_i}{\partial t}+\left(v^2-4U_0\right)\frac{\partial U_1}{\partial x^i}-4v_iv_j\frac{\partial U_1}{\partial x^j}-4U_1\frac{\partial U_0}{\partial x^i}
\\\lb{jpn7b}
-3v_i\frac{\partial U_1}{\partial t}+v_j\left(\frac{\partial \Pi^1_i}{\partial x^j}-\frac{\partial \Pi^1_j}{\partial x^i}\right)\Bigg]\frac{\partial f_0}{\partial v^i}=0.\qquad
\een
In the above equations we have written the Maxwell-J\"uttner distribution function as
\ben\lb{jpn7c}
f_{MJ}=f_0
\left\{1-\frac{\sigma^2}{c^2}\left[\frac{15}8+\frac{3v^4}{8\sigma^4}
+2\frac{U_0 v^2}{\sigma^4}\right]\right\}-2f_0\epsilon\frac{U_1v^2}{c^2\sigma^2}
=f_{MJ}^0-2f_0\epsilon\frac{U_1v^2}{c^2\sigma^2},
\een
where the background Maxwell-J\"uttner distribution function was denoted by $f_{MJ}^0$.

We note that the background terms are related with  a stationary equilibrium state so that the background equation (\ref{jpn7a}) becomes an identity when the gradients of the  gravitational potential backgrounds  vanish, i.e., $\nabla U_0=0$, $\nabla\Phi_0=0$ and $\nabla \Pi_i^0=0$. By considering that the gravitational potential backgrounds are constants the perturbed Boltzmann equation (\ref{jpn7b})  reduces to
\ben\no
\frac{\partial f_1}{\partial t}+v_i\frac{\partial f_1}{\partial x^i}+\frac{\partial U_1}{\partial x^i}\frac{\partial f_{MJ}^0}{\partial v^i}-\frac{2v^2f_0}{\sigma^2c^2}\left(\frac{\partial U_1}{\partial t}+v_i\frac{\partial U_1}{\partial x^i}\right)
+\frac1{c^2}\Bigg[\left(v^2-4U_0\right)\frac{\partial U_1}{\partial x^i}+2\frac{\partial \Phi_1}{\partial x^i}+\frac{\partial \Pi^1_i}{\partial t}-3v_i\frac{\partial U_1}{\partial t}
\\\lb{jpn8}
-4v_iv_j\frac{\partial U_1}{\partial x^j}+v_j\left(\frac{\partial \Pi^1_i}{\partial x^j}-\frac{\partial \Pi^1_j}{\partial x^i}\right)\Bigg]\frac{\partial f_0}{\partial v^i}=0.
\een

One can observe from the Poisson equations (\ref{jpn2a}) and (\ref{jpn2b}) that they are not  satisfied by the conditions of vanishing background potential gravitational gradients, since the right-hand sides of (\ref{jpn2a}) and (\ref{jpn2b})  are given in terms of the energy-momentum tensor which does not vanish at equilibrium. At this point we assume  "Jeans  swindle" (see e.g. \cite{BT1}) to remove this inconsistency    and consider that the Poisson equations are valid only for the perturbed distribution function and gravitational potentials.

We note from the perturbed Boltzmann equation (\ref{jpn8}) that it is a function of the background value of the Newtonian gravitational potential $U_0$ which is a constant. In the analysis of the Jeans instability based on the post-Newtonian hydrodynamic equations \cite{NKR} it was supposed vanishing values for the  background gravitational potentials as a part of the "Jeans  swindle". Here we shall not adopt this statement and will show  that the background Newtonian gravitational potential $U_0$  has a prominent role in the determination of Jeans mass.

For the determination of the  energy-momentum tensor components (\ref{jpn3a})  we have to write the  the four-velocity components (\ref{jpn3b}) and the invariant integration element (\ref{jpn4b}) by taking into account the representation of the Newtonian gravitational potential (\ref{jpn6b}), yielding
\ben\lb{jpn9a}
u^0=c\left[1+\frac1{c^2}\left(\frac{v^2}2+U_0+\epsilon U_1\right)\right],\qquad u^i=\frac{u^0v^i}c,
\\\lb{jpn9b}
\frac{\sqrt{-g}\, d^3 u}{u_0}=
\left\{1+\frac1{c^2}\left[2v^2+6U_0+6\epsilon U_1\right]\right\}\frac{d^3v}c.
\een

We multiply the one-particle distribution function    (\ref{jpn6a}) together with (\ref{jpn7c}) with the invariant element  (\ref{jpn9b}) and get
\ben\no
f\frac{\sqrt{-g}\, d^3 u}{u_0}=\left\{1-\frac{\sigma^2}{c^2}\left[\frac{15}8+\frac{3v^4}{8\sigma^4}+2\frac{U_0v^2}{\sigma^4}-2\frac{v^2}{\sigma^2}\-6\frac{U_0}{\sigma^2}\right]\right\}f_0\frac{d^3v}c
-\epsilon\frac{U_1}{c^2}\left(\frac{2v^2}{\sigma^2}-6\right)f_0\frac{d^3v}c
\\\lb{jpn9c}
+\epsilon\left\{1+\frac{1}{c^2}\left[2v^2+6U_0\right]\right\}f_1\frac{d^3v}c.
\een

Now we can evaluate the  energy-momentum tensor components that appear in the right-hand sides of the Poisson equations (\ref{jpn2a}) and (\ref{jpn2b}), by inserting the expressions  (\ref{jpn9a}) and (\ref{jpn9c}) into the definition of the energy-momentum tensor (\ref{jpn3a}), resulting
\ben\no
&&{\buildrel\!\!\!\! _{0} \over{T^{00}}}+{\buildrel\!\!\!\! _{2} \over{T^{00}}}=m^4c\int u^0 u^0 f\frac{\sqrt{-g}\,d^3 u}{u_0}
=m^4c^2\int f_0\Bigg[1-\frac{\sigma^2}{c^2}\Bigg(\frac{15}8+\frac{3v^4}{8\sigma^4}+\frac{2U_0v^2}{\sigma^4}-\frac{3v^2}{\sigma^2}-\frac{8U_0}{\sigma^2}\Bigg)\Bigg]d^3v
\\\lb{jpn10a}
&&\qquad+\epsilon\,m^4c^2\int\Bigg\{ f_1 \Bigg[1+\frac{3v^2}{c^2}+\frac{8U_0}{c^2}\Bigg]
-\Bigg(\frac{2v^2}{\sigma^2}-8\Bigg)\frac{f_0U_1}{c^2}\Bigg\}d^3v,
\\\lb{jpn10b}
&&{\buildrel\!\!\!\! _{2} \over{T^{ij}}}=m^4c\int u^i u^j f\frac{\sqrt{-g}\,d^3 u}{u_0}=m^4\int v_iv_j(f_0+\epsilon f_1)d^3v,
\\\lb{11c}\lb{jpn10c}
&& {\buildrel\!\!\!\! _{1} \over{T^{0i}}}=m^4c\int u^0 u^i f\frac{\sqrt{-g}\,d^3 u}{u_0}=m^4c\int v_i(f_0+\epsilon f_1)d^3v.
\een

By taking into account the expressions (\ref{jpn10a}) -- (\ref{jpn10c}) for  the energy-momentum tensor components we get that  the perturbed Poisson equations  (\ref{jpn2a}) and (\ref{jpn2b}) become
\ben\lb{jpn11a}
&&\nabla^2U_1=-\frac{4\pi G}{c^2} [{\buildrel\!\!\!\! _{0} \over{T^{00}}}]_1=-4\pi G m^4\int f_1 d^3v,
\\\lb{jpn11b}
&&\nabla^2\Pi_1^i=-\frac{16\pi G}{c}[{\buildrel\!\!\!\! _{1} \over{T^{0i}}}]_1+\frac{\partial^2U_1}{\partial t\partial x^i}
=-16\pi G m^4\int v_if_1d^3v+\frac{\partial^2U_1}{\partial t\partial x^i},
\\\lb{jpn11c}
&& \nabla^2\Phi_1=-2\pi G \left([{\buildrel\!\!\!\! _{2} \over{T^{00}}}]_1+[{\buildrel\!\!\!\! _{2} \over{T^{ii}}}]_1\right)=
- 2\pi G m^4\int\bigg[\left(4v^2
 +8U_0\right)f_1 -\left(\frac{2v^2}{\sigma^2}-8\right)U_1f_0\bigg]d^3v.
\een
In the above equations the  energy-momentum tensor components calculated with the perturbed distribution function $f_1$ were denoted by  $[{\buildrel\!\!\!\! _{0} \over{T^{00}}}]_1$, $[{\buildrel\!\!\!\! _{1} \over{T^{0i}}}]_1$ and so one. 

As usual for the search of the instabilities the perturbations are represented as plane waves of frequency $\omega$ and wave number vector $\bk$, namely
\ben\lb{jpn12a}
f_1(\bx,\bv,t)=\overline f_1e^{i(\bk\cdot\bx-\omega t)},\quad U_1(\bx,\bv,t)=\overline U_1e^{i(\bk\cdot\bx-\omega t)},
\\\lb{jpn12b}
\Phi_1(\bx,\bv,t)=\overline \Phi_1e^{i(\bk\cdot\bx-\omega t)},\quad \Pi^i_1(\bx,\bv,t)=\overline {\Pi_i^1}e^{i(\bk\cdot\bx-\omega t)},
\een
where $\overline f_1, \overline U_1,\overline\Phi_1$ and $\overline{\Pi_1^i}$ represent small amplitudes of the perturbations.

If we  insert the plane wave representations (\ref{jpn12a}) and (\ref{jpn12b}) into the perturbed Boltzmann equation (\ref{jpn8}) we get
\ben\no
({\bf v\cdot k}-\omega)\overline f_1-\frac{f_0}{\sigma^2}\bigg\{({\bf v\cdot k})\overline U_1\bigg[1-\frac{\sigma^2}{c^2}\bigg(\frac{15}8+\frac{3v^4}{8\sigma^4}-\frac{v^2}{2\sigma^2}
+\frac{2v^2U_0}{\sigma^4}\bigg)\bigg]+\frac1{c^2}\left[v^2\omega\overline U_1+2({\bf v\cdot k})\overline\Phi_1-\omega v_i\overline{\Pi^1_i}\right]\bigg\}=0,
\\\lb{jpn13a}
\een
by taking into account the expression (\ref{jpn7c}) for the determination of the term $\partial f_{MJ}/\partial v^i$.

The  Poisson  equations (\ref{jpn11a}) -- (\ref{jpn11c}) with the  plane wave representations (\ref{jpn12a}) and (\ref{jpn12b}) become
\ben\lb{jpn13b}
&&\kappa^2\overline U_1=4\pi G m^4\int \overline f_1d^3v,
\\\lb{jpn13c}
&&\kappa^2\overline{\Pi^1_i}=16\pi G m^4\int v_i\overline f_1d^3v+k_i\omega\overline U_1,
\\\lb{jpn13d}
&&\kappa^2\overline\Phi_1=8\pi G m^4\int (v^2+2U_0)\overline f_1d^3v+4\pi G\rho_0\overline U_1.
\een

 We have to evaluate the integrals in (\ref{jpn13b}) -- (\ref{jpn13d}) and for that end we choose, without loss of generality,  the wave number vector in the $x$-direction, i.e.,  ${\bf k}=(\kappa,0,0)$. We begin  with  the substitution of $\overline f_1$ from (\ref{jpn13a}) into (\ref{jpn13c}) and get
\ben\no
\kappa^2\overline{\Pi^1_i}=\frac{16\pi G \rho_0}{(2\pi\sigma^2)^\frac32}\int \frac{v_i(v_x\kappa+\omega)e^{-\frac{v^2}{2\sigma^2}}d^3v}{\sigma^2[(v_x \kappa)^2-\omega^2]}
\bigg\{ v_x \kappa \bigg[1-\frac{\sigma^2}{c^2}\bigg(\frac{15}8+\frac{3v^4}{8\sigma^4}-\frac{v^2}{2\sigma^2}+\frac{2U_0v^2}{\sigma^4}\bigg)\bigg]\overline U_1
\\\lb{jpn14a}
+\frac1{c^2}\bigg[v^2\omega\overline U_1+2v_x \kappa \overline\Phi_1-\omega v_j \overline{\Pi^1_j}\bigg]\bigg\}+k_i\omega\overline U_1.
\een
Here the numerator and denominator of the integrand were multiplied by $(v_x\kappa+\omega)$.

 For the components $i=y,z$ the integration of (\ref{jpn14a})  in the ranges $-\infty<(v_x,v_y,v_z)<\infty$ leads to
\ben\lb{jpn15}
\kappa^2\overline{\Pi^1_i}=-8\pi G\rho_0\frac{\omega^2}{\kappa^2\sigma^2c^2}I_0\overline{\Pi^1_i},\qquad i=y,z,
\een
where $I_0$ is an integral which is given in terms of the  $I_n$ integrals defined by
\be
I_n(\kappa,\omega)=\frac2{\sqrt\pi}\int_0^\infty\frac{x^ne^{-x^2}}{x^2-(\omega/\sqrt2\sigma \kappa)^2}dx,\quad
x=\frac{v_x}{\sqrt2\sigma}.
\ee{jpn16}

We conclude from (\ref{jpn15})  that $\overline{\Pi^1_y}=\overline{\Pi^1_z}=0$.

 For the component $i=x$ the integration of (\ref{jpn14a})  in the ranges $-\infty<(v_x,v_y,v_z)<\infty$, yields
\ben\lb{jpn17a}
\kappa^2\overline{\Pi_x^1}=\kappa\omega\overline U_1+\frac{16\pi G\rho_0\omega}{\kappa\sigma^2}\bigg\{\bigg[I_2-\frac{3\sigma^2}{2c^2}\left(I_6+\frac54I_2\right)
-\frac{4U_0}{c^2}\left(I_2+I_4\right)\bigg]\overline U_1+\frac{I_2}{c^2}\left[2\overline\Phi_1-\frac{\omega}{\kappa}\overline{\Pi_x^1}\bigg]\right\}.
\een

Now we follow  the same methodology and substitute  $\overline f_1$ from (\ref{jpn13a}) into  (\ref{jpn13b}) and (\ref{jpn13d}). The subsequent integration of  the resulting equations  in the ranges $-\infty<(v_x,v_y,v_z)<\infty$ lead to
\ben\lb{jpn17b}
\kappa^2\overline U_1 =\frac{4\pi G\rho_0}{\sigma^2}\bigg\{\bigg[I_2+\left(I_0+I_2\right)\frac{\omega^2}{c^2\kappa^2}
-\frac{3\sigma^2}{2c^2}\bigg(I_6+\frac43I_4+\frac{31}{12}I_2\bigg)-\frac{4U_0}{c^2}\left(I_2+I_4\right)\bigg]\overline U_1
+\frac{I_2}{c^2}\bigg[2\overline\Phi_1-\frac{\omega}{\kappa}\overline{\Pi_x^1}\bigg]\bigg\},\quad
\\\no
\kappa^2\overline\Phi_1 =4\pi G\rho_0\overline U_1+16\pi G\rho_0\bigg\{2\bigg(I_0+I_2+\frac{I_4}2\bigg)\frac{\omega^2}{\kappa^2c^2}
+I_2+I_4-\frac{3\sigma^2}{2c^2}\bigg(I_8+\frac73I_6+\frac{71}{12}I_4+\frac{71}{12}I_2\bigg)+\frac{U_0}{\sigma^2}\bigg[I_2
\\\lb{jpn17c}
-\frac{\sigma^2}{c^2}\bigg(\frac{11}2I_6+10I_4+\frac{95}8I_2\bigg)
-\frac{4U_0}{c^2}(I_2+I_4)+\frac{\omega^2}{c^2\kappa^2}(I_0+I_2)\bigg]\bigg\}\overline U_1
+\frac{16\pi G\rho_0}{c^2}\left(I_2+I_4+I_2\frac{U_0}{\sigma^2}\right)\left[2\overline\Phi_1-\frac{\omega}{\kappa}\overline{\Pi_x^1}\right].\quad
\een

By inspecting equations  (\ref{jpn17a}) --  (\ref{jpn17c}) we conclude that they represent  an algebraic system of equations for the amplitudes $\overline{\Pi_x^1}$, $\overline U_1$ and $\overline\Phi_1$. This system of equations admits a solution  if the determinant of the coefficients of $\overline{\Pi_x^1}$, $\overline U_1$ and $\overline\Phi_1$ vanishes. Hence we get the following   dispersion relation
\ben\lb{jpn18a}
\kappa_*^4-\kappa_*^2\left[I_2+\frac{\sigma^2}{c^2}\left(\frac{33}8I_2+6I_4-\frac32I_6+4(I_2-I_4)\frac{U_0}{\sigma^2}\right)\right]
-\frac{\sigma^2}{c^2}\Bigg[2I_2+\left(I_0-4I_2\right)\omega_*^2\Bigg]=0.
\een
The  above dispersion relation is an algebraic equation which relates the dimensionless wave number $\kappa_*$ with the dimensionless frequency $\omega_*$. They are defined by
\ben\lb{jpn18b}
\kappa_*=\frac{\kappa}{\kappa_J},\qquad \omega_*=\frac\omega{\sqrt{4\pi G\rho_0}},
\een
where $\kappa_J=\sqrt{4\pi G\rho_0}/\sigma$
denotes the Jeans wave number. 

Note that in the  dispersion relation (\ref{jpn18a}) we have not considered the terms that have  order higher than  $1/c^{2}$, due to the fact that  we are considering only the first post-Newtonian approximation.

The perturbations will propagate as  harmonic waves in time if the frequency $\omega$ has real values, while for pure imaginary values of the frequency  the perturbation will grow or decay in time. The one which grows with time is associated with the Jeans instability. Hence, the corresponding solutions to the Jeans instability are those where $\omega_*=i\omega_I$, i.e., $\Re(\omega_*)=0$ and $\omega_I=\Im(\omega_*)>0$. The   integrals  (\ref{jpn16}) in this case can be evaluated, yielding
\ben\lb{jpn19a}
I_0=\frac{\kappa_*}{\omega_I}\sqrt{2\pi}\exp\left(\frac{\omega_I^2}{2\kappa_*^2}\right){\rm erfc}\left(\frac{\omega_I}{\sqrt2\kappa_*}\right),\quad
I_2=1-\frac{\omega_I^2}{2\kappa_*^2}I_0,\quad I_4=\frac12-\frac{\omega_I^2}{2\kappa_*^2}I_2,\quad I_6=\frac34-\frac{\omega_I^2}{2\kappa_*^2}I_4.
\een
Here $\rm{erfc}(x)$ is the complementary error function
\ben\lb{jpn19c}
\rm{erfc}(x)=\frac2{\sqrt\pi}\int_x^\infty e^{-x^2}dx.
\een

\begin{figure}[ht]\begin{center}\includegraphics[width=7.5cm]{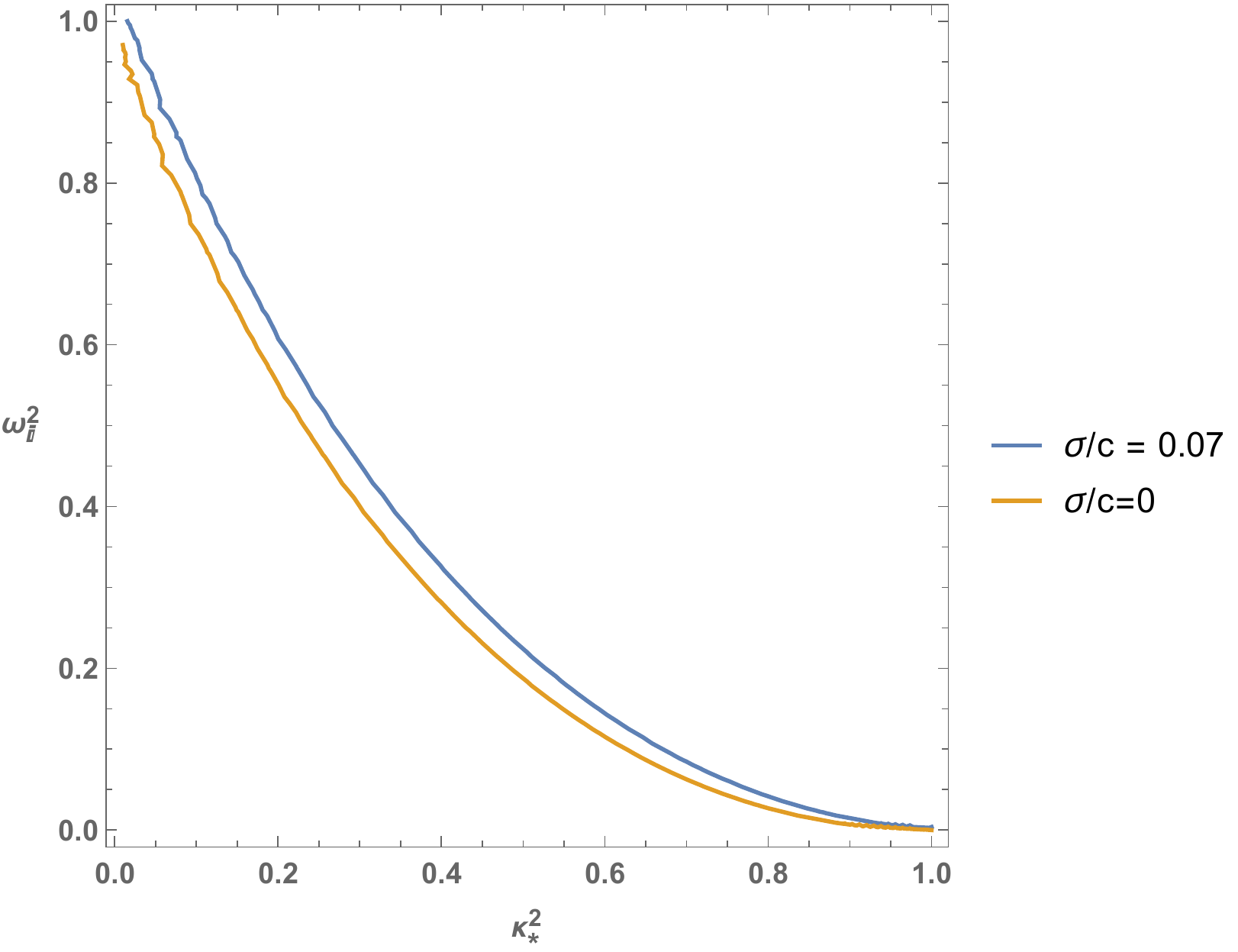}\end{center}\caption{Dimensionless frequency  as function of the dimensionless wave number vector  modulus for the post-Newtonian  ($\sigma/c=0.07$) and Newtonian ($\sigma/c=0$)  theories.}\lb{fig1}\end{figure}

The contour plots which follow from  the dispersion relation (\ref{jpn18a}) are shown in Figure \ref{fig1} for two different values of the ratio between the dispersion velocity and the light speed, namely $\sigma/c=0$ which corresponds to the Newtonian theory and $\sigma/c=0.07$ to the post-Newtonian theory. For the evaluation of (\ref{jpn18a}) it was assumed  that $U_0\approx\sigma^2$, which can be justified by the virial theorem where the square of the dispersion velocity can be approximated with the Newtonian gravitational potential.
We observe from this figure that the limit of instability in the post-Newtonian theory differs from the one of the  Newtonian theory. Indeed,  the modulus of the wave number vector for a given frequency  in the Newtonian theory is smaller than that of the post-Newtonian theory. 
As a consequence, the mass limit of interstellar gas clouds necessary to start the gravitational collapse in the post-Newtonian theory is smaller than the one in the Newtonian theory.  

Let us investigate   the limiting value of the frequency where the instability occurs and which corresponds to the  minimum mass where an overdensity begins the gravitational collapse. For that end we set $\omega_I=0$ in (\ref{jpn18a}), yielding
\ben\lb{jpn20a}
\kappa_*^4-\left[1+\frac{\sigma^2}{c^2}\left(6+\frac{2U_0}{\sigma^2}+\frac2{\kappa_*^2}\right)\right]\kappa_*^2=0.
\een
The   solution of the fourth order algebraic equation (\ref{jpn20a}) is
\ben\lb{jpn20b}
\kappa_*=\pm\sqrt{\frac12+\frac{\sigma^2}{c^2}\left[3+\frac{U_0}{\sigma^2}\pm\sqrt{\frac14+\frac{\sigma^2}{c^2}\left[5+\frac{U_0}{\sigma^2}+\frac{\sigma^2}{c^2}\left(9+\frac{6U_0}{\sigma^2}+\frac{U_0^2}{\sigma^4}\right)\right]}\right]}
\een

The real positive value of $\kappa_*$  when terms up to the $1/c^2$ order are considered reads
\ben\lb{jpn20c}
\kappa_*=1+\frac{\sigma^2}{c^2}\left[4+\frac{U_0}{\sigma^2}\right].
\een

Here we shall call attention to the fact that the post-Newtonian correction is given in terms of the ratio of the dispersion velocity $\sigma$ and the light speed $c$. In a phenomenological theory this correction is given in terms of the adiabatic sound speed $c_s$ and the light speed $c$. This difference comes out that the Maxwellian distribution function is written in terms of the dispersion velocity while in the phenomenological theory an adiabatic solution is considered. 

In a recent paper Noh and Hwang \cite{NH} obtained  from a dispersion relation the post-Newtonian correction which  corresponds to (\ref{jpn20c}). In our notation  the real root of equation (78) of \cite{NH} in the post-Newtonian approximation reads
\ben
\kappa_*=1+\frac{c_s^2}{c^2}\left[\frac{\left(\Pi_0+\frac{p_0}{\rho_0}\right)}{c_s^2}+\frac52+\frac{U_0}{c_s^2}\right].
\een
Here $\Pi_0=(p_0/\rho_0)/(\gamma-1)$ is the specific internal energy. By considering the adiabatic sound velocity $c_s^2=\gamma p_0/\rho_0$ and $\gamma=5/3$ the above equation reduces to
\ben
\kappa_*=1+\frac{c_s^2}{c^2}\left[4+\frac{U_0}{c_s^2}\right],
\een
which has the same structure as (\ref{jpn20c}). This result from a phenomenological theory is the same as the one found in \cite{GGKK}.

 The Jeans  mass is related with the minimum amount  of  mass  for an overdensity  to  initiate  the  gravitational collapse and refers to the mass contained in a sphere of radius equal to the wavelength of the perturbation. If we denote the mass corresponding to the post-Newtonian wavelength by $M_J^{PN}$ and the Newtonian one by $M_J^N$ wavelengths, their ratio is given by 
\ben\lb{jpn20d}
\frac{M_J^{PN}}{M_J^N}=\frac{\lambda^3}{\lambda_J^3}=\frac{\kappa_J^3}{\kappa_*^3}\approx 1-3\frac{\sigma^2}{c^2}\left[4+\frac{U_0}{\sigma^2}\right].
\een
From the above equation we infer that in the post-Newtonian framework the mass needed to begin the gravitational collapse is smaller than in the Newtonian one. Furthermore, we note that the background Newtonian potential has an important role, since it implies a smaller mass than the one without it. As was previously commented one can make use of the virial theorem to approximate $U_0\approx \sigma^2$, so that (\ref{jpn20d}) becomes
\ben\lb{jpn20d}
\frac{M_J^{PN}}{M_J^N}= 1-15\frac{\sigma^2}{c^2}.
\een

\section{Summary}\lb{sec4}

In this work the Jeans instability was analysed within the framework of the Boltzmann and Poisson equations that follow from the first post-Newtonian theory. The components of the energy-momentum tensor in the Poisson equations were determined from the Maxwell-J\"uttner distribution function. The distribution function and the gravitational potentials were perturbed from their background states and a plane wave representation for the perturbations was considered. The post-Newtonian dispersion relation was obtained from an algebraic system of equations for the perturbed gravitational potentials. It was shown that the
  mass necessary for an overdensity to begin the gravitational collapse in the post-Newtonian theory is smaller than the one in the Newtonian one. Furthermore,  a non-vanishing  Newtonian gravitational potential background implies a smaller Jeans mass than the one where  a vanishing value is considered \cite{NKR}.

\section*{Acknowledgments} This work was supported by Conselho Nacional de Desenvolvimento Cient\'{i}fico e Tecnol\'{o}gico (CNPq), grant No. \break 304054/2019-4.

\end{document}